\documentclass{iopart}
\usepackage{graphicx}

\newcommand \beq{\begin{eqnarray}}
\newcommand \eeq{\end{eqnarray}}
\newcommand{\mnote}[1]{\marginpar{\tiny {}}}   

\def\rez{$^{(1)}$}
\def\gsi{$^{(2)}$}
\def\hei{$^{(3)}$}
\def\dub{$^{(4)}$}
\def\wei{$^{(5)}$}
\def\sun{$^{(6)}$}
\def\cer{$^{(7)}$}
\def\bnl{$^{(8)}$}
\def\mpi{$^{(9)}$}

\begin{document}

\title{Lambda production in 40 A\,GeV/c Pb-Au collisions}
\author{Wolfgang Schmitz for the CERES Collaboration, D.~Adamov\'a\rez, 
G.~Agakichiev\gsi, 
H.~Appelsh\"auser\hei, 
V.~Belaga\dub, 
P.~Braun-Munzinger\gsi, 
A.~Cherlin\wei, 
S.~Damjanovi\'c\hei, 
T.~Dietel\hei, 
L.~Dietrich\hei, 
A.~Drees\sun, 
S.\,I.~Esumi\hei, 
K.~Filimonov\hei, 
K.~Fomenko\dub,
Z.~Fraenkel\wei, 
C.~Garabatos\gsi, 
P.~Gl\"assel\hei, 
G.~Hering\gsi, 
V.~Kushpil\rez, 
B.~Lenkeit\cer, 
A.~Maas\gsi, 
A.~Mar\'{\i}n\gsi, 
F.~Messer\sun, 
J.~Milo\v{s}evi\'c\hei,
A.~Milov\wei, 
D.~Mi\'skowiec\gsi, 
Yu.~Panebrattsev\dub, 
O.~Petchenova\dub, 
V.~Petr\'a\v{c}ek\hei, 
A.~Pfeiffer\cer, 
J.~Rak\gsi, 
I.~Ravinovich\wei, 
P.~Rehak\bnl, 
H.~Sako\gsi, 
W.~Schmitz\hei, 
J.~Schukraft\cer, 
S.~Sedykh\gsi, 
S.~Shimansky\dub, 
J.~Sl\'{\i}vov\'a\hei,
H.\,J.~Specht\hei, 
J.~Stachel\hei, 
M.~\v{S}umbera\rez, 
H.~Tilsner\hei, 
I.~Tserruya\wei, 
J.\,P.~Wessels\hei, 
T.~Wienold\hei, 
J.\,P.~Wurm\mpi, 
W.~Xie\wei, 
S.~Yurevich\hei, 
V.~Yurevich\dub}

\address{
\rez NPI ASCR, \v{R}e\v{z}, Czech Republic\\
\gsi GSI Darmstadt, Germany\\
\hei Heidelberg University, Germany\\
\dub JINR Dubna, Russia\\
\wei Weizmann Institute, Rehovot, Israel\\
\sun SUNY at Stony Brook, U.S.A.\\
\cer CERN, Geneva, Switzerland\\
\bnl BNL, Upton, U.S.A.\\
\mpi MPI, Heidelberg, Germany\\}

\begin{abstract}
  During the 1999 lead run, CERES has measured hadron and
  electron-pair production at 40 A\,GeV/c beam momentum with the
  spectrometer upgraded by the addition of a radial TPC. Here the
  analysis of $\Lambda$ and $\bar\Lambda$ will be presented.  
\end{abstract}

\section{Introduction}

\begin{figure}[tbh]
  \begin{center}
    \includegraphics[height=125mm,angle=-90]{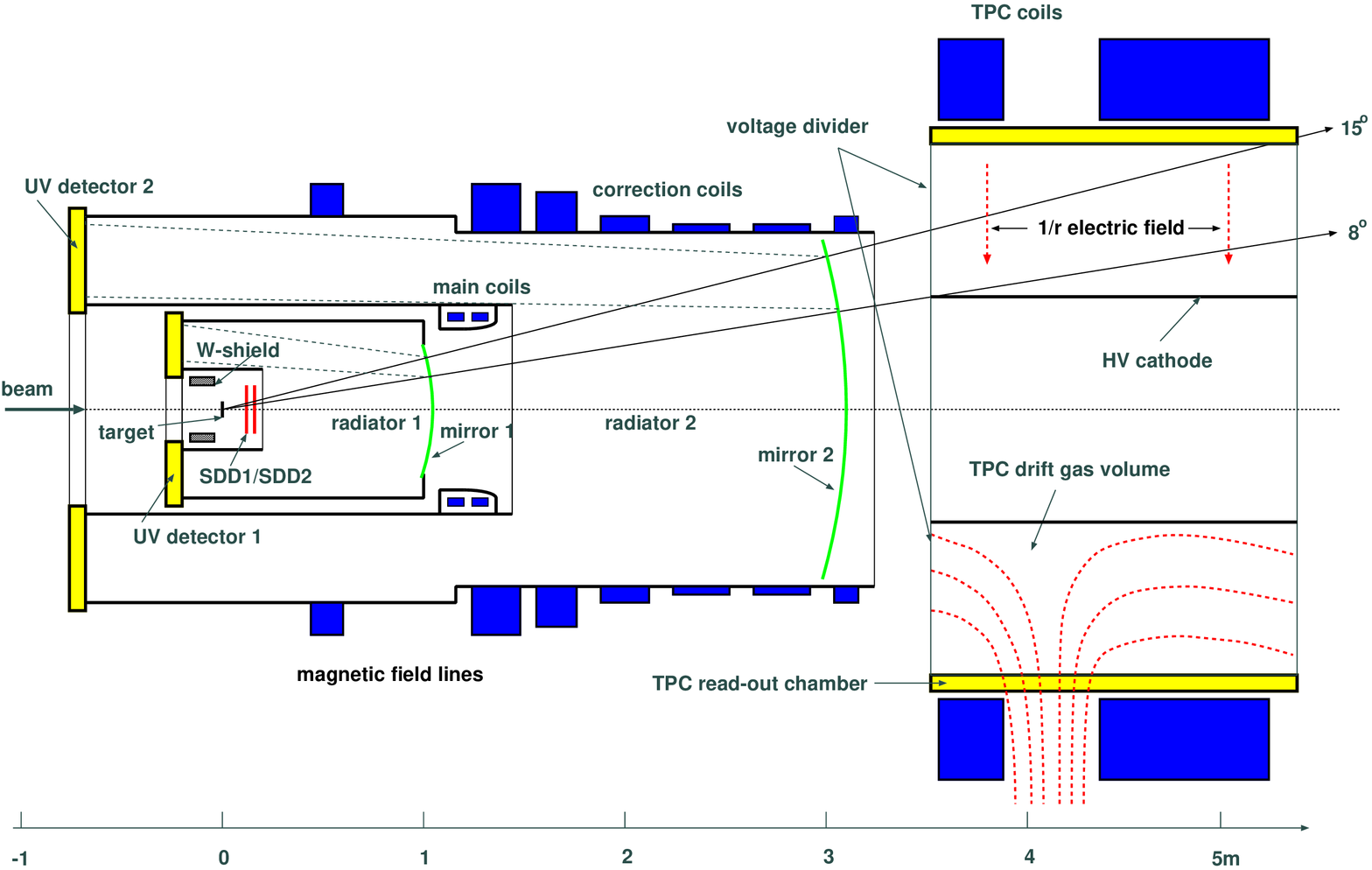}
  \end{center}
  \caption{The upgraded CERES spectrometer at the CERN SPS. In
    addition to the outlines of the various detectors are shown in the
    TPC field lines for the electric and magnetic field.}
\label{tpc_trunc_field}
\end{figure}

CERES/NA45 is an experiment at the CERN SPS dedicated to measure
low-mass e$^+$e$^-$ pairs near midrapidity in ultrarelativistic
nuclear collisions \cite{oldceres}. The CERES spectrometer
(Fig.~\ref{tpc_trunc_field}) covers the full azimuthal acceptance in the
polar angle region between $8^ {\circ}$ $<$ $\theta$ $<$ $14.6^
{\circ}$. The high electron identification capability allows to separate
the rare leptonic signals from the large hadronic background; it is
provided by two ring-imaging Cerenkov counters (RICH). Precise tracking
of charged particles and vertex reconstruction are provided
by two silicon drift detectors (SDD1,2) located behind a segmented Au
target. This detector system is also used as a multiplicity trigger to
get information on the centrality of the collision.

In order to improve the momentum resolution the CERES experiment has
been upgraded \cite{TechNot,Marin} during 1998 by the addition of a new
magnet system and a Time Projection Chamber (TPC) with a radial electric
drift field. The TPC is operated inside an inhomogeneous magnetic field with
a maximal radial component of 0.5 T, and provides the measurement of up to
twenty space points for each charged particle track. Besides a precise
determination of the momentum the TPC also provides additional
electron identification via dE/dx. The magnet system between the two
RICHes was not operated in the upgraded configuration.

The addition of the TPC substantially improves the hadron capability
of the CERES spectrometer allowing a systematic investigation of
hadronic observables around midrapity.  In the fall of 1999 the
upgraded CERES spectrometer was operated for the first time and 
$8\cdot 10^6$ semicentral Pb+Au collisions at 40 A\,GeV/c beam momentum were
recorded. Because of a not yet completely functional read-out system
this data set is limited in terms of statistics and momentum
resolution. In the fall of 2000, the CERES
spectrometer was fully operational with a very good overall performance,
and $33\cdot 10^6$ central 158 A\,GeV/c Pb+Au events were recorded. In this
paper we will focus on the $\Lambda$ analysis of the 40 A\,GeV/c data set.

\section{Analysis of the $\Lambda$ hyperon signal}

\begin{figure}[tbh]
    \includegraphics[width=4.4cm]{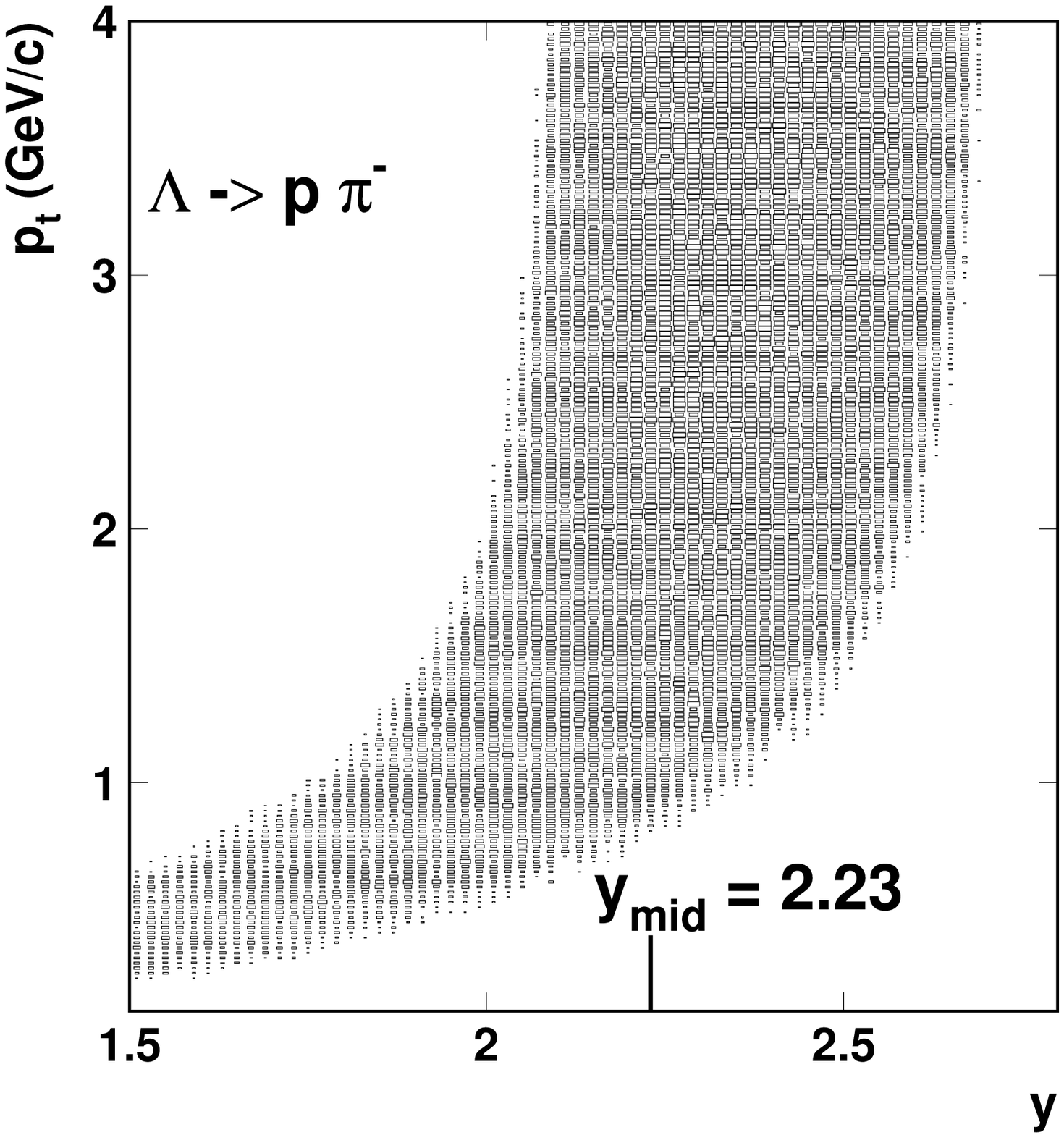}
    \includegraphics[width=4.4cm]{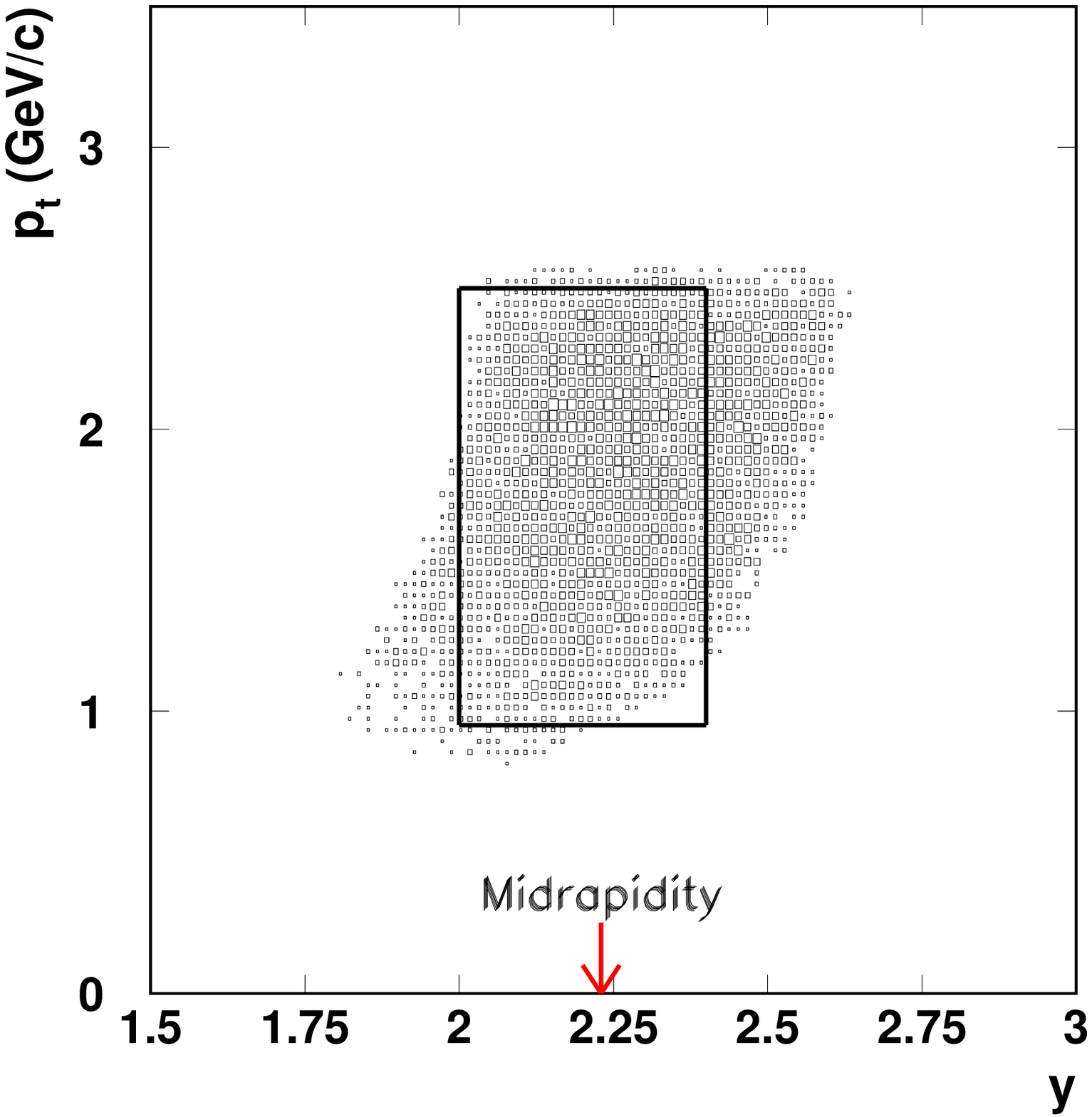}
    \includegraphics[width=4.4cm]{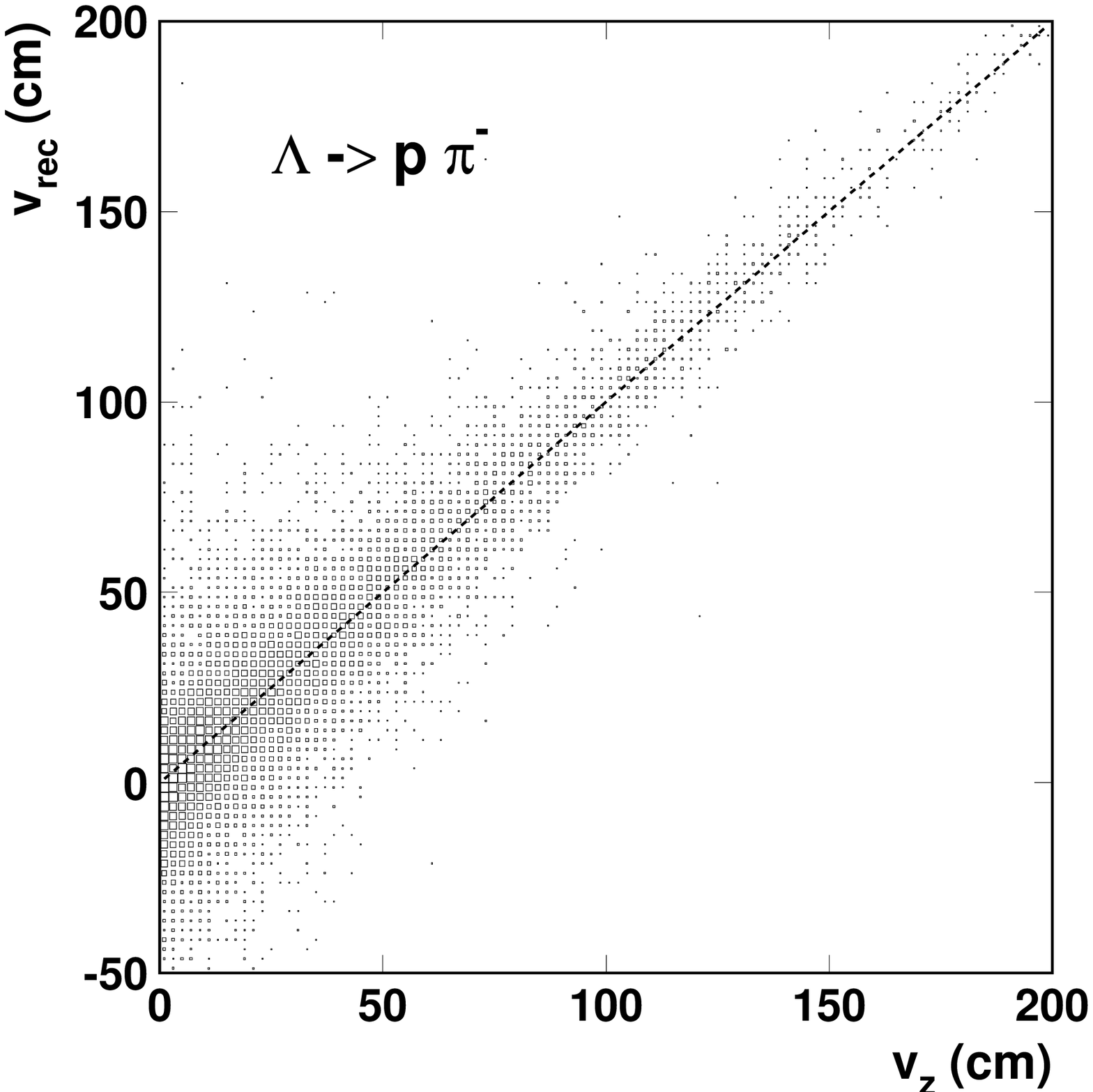}
     \caption{Simulated geometrical acceptance of the TPC for
    $\Lambda$ hyperons decaying into a proton and a pion. Left: no cuts,
    Middle: all cuts applied (see text). The box indicates the region
    selected by pair cuts. Right: Reconstructed decay point $v_{rec}$
    vs true decay point $v_z$ in beam direction from a GEANT simulation.}
    \label{simulation}
\end{figure}

In this $\Lambda$ analysis we have used only the information from the SDD's
and the TPC. The $\Lambda$ hyperon and its antiparticle were
identified by reconstructing their decays into final states containing
only charged particles: $\Lambda ~ \rightarrow ~ \rm p ~ \pi^-$ and
$\bar{\Lambda} ~ \rightarrow ~ \bar{\rm p} ~ \pi^+$. The geometrical
acceptance for the $\Lambda$ measurement is shown on the left side in
Fig.~\ref{simulation}. We have also analyzed negatively charged
particles (h$^-$) and proton-like positive net charges, shortly denoted
as $(+)-(-)$ \cite{Harry}.

The momentum of charged particles is determined from their
curvature in the TPC. Since we are not applying any particle
identification method in the extraction of the strange particle
signals we run through the combinations of all positive tracks with
all negative tracks measured in the TPC assuming a decay
hypothesis and calculate the invariant mass spectrum. In the same
manner, the combinatorial background is evaluated using the mixed
event method.

To get a significant $\Lambda$ signal we had to reduce the large 
combinatorial background using different types of cuts discussed in
the following. 

{\bf Discrimination from target tracks}: Taking into account the proper
decay length $c\tau_0$ = 7.89 cm of the $\Lambda$ and the
$\gamma$-factor of about 5 for the TPC acceptance, the decay length in
the laboratory is $\beta c\tau_{lab} = \beta\gamma c\tau_0 \approx$ 40
cm. This means that more than 70 \% of all $\Lambda$ produced in the
target decay after the SDD's (the distance between target and SDD2
is about 13 cm). To enhance the signal-to-background ratio, an
optional cut requiring the absence of matching tracks in the SDD is
used.

{\bf Reconstruction of secondary vertex:} In a second step, the decay
vertex $v_{rec}$ in z-direction (beam direction) is reconstructed for
all $\Lambda$ candidates by back-extrapolation of the TPC tracks and
computing the point of their closest approach. This procedure was
tested using a simulation. The result is shown in
Fig.~\ref{simulation}, right. This reconstructed secondary vertex is
used in the analysis for an additional (optional) cut $v_{rec} \geq
50$ cm. At this threshold, the resolution in $v_{rec}$ is 12 cm.

\begin{figure}[tbh]
\vspace*{3mm}
  \includegraphics[width=61mm,bbllx=10,bblly=40,bburx=510,bbury=510]{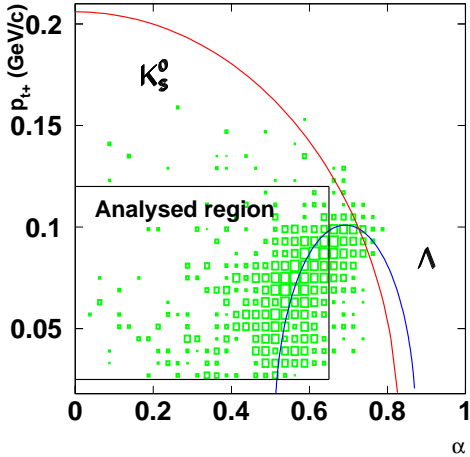}

\caption{Armenteros diagram. }
\label{armenteros}
\end{figure}

\begin{figure}[tbh]
  \includegraphics[width=67mm]{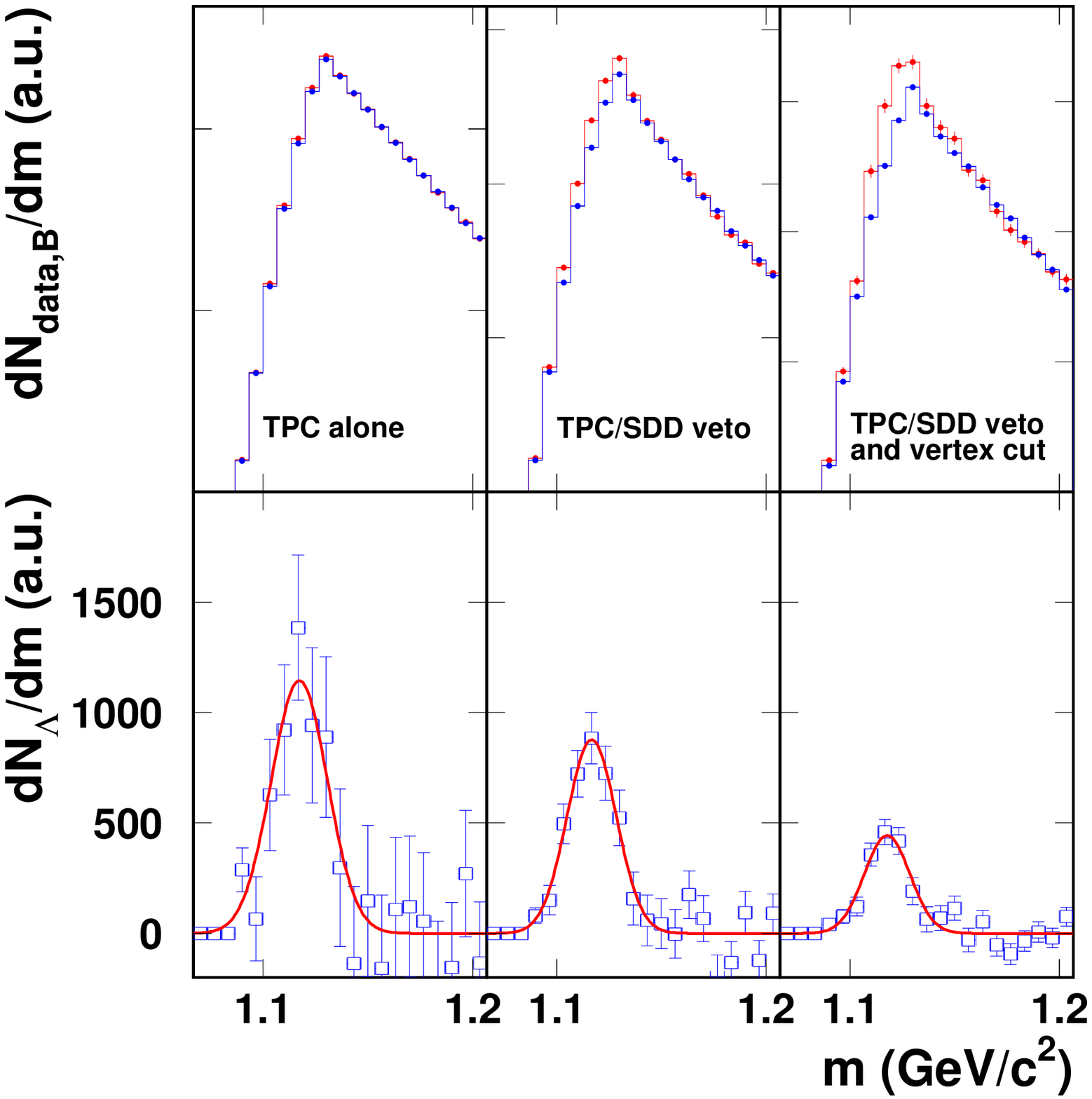}
  \includegraphics[width=67mm]{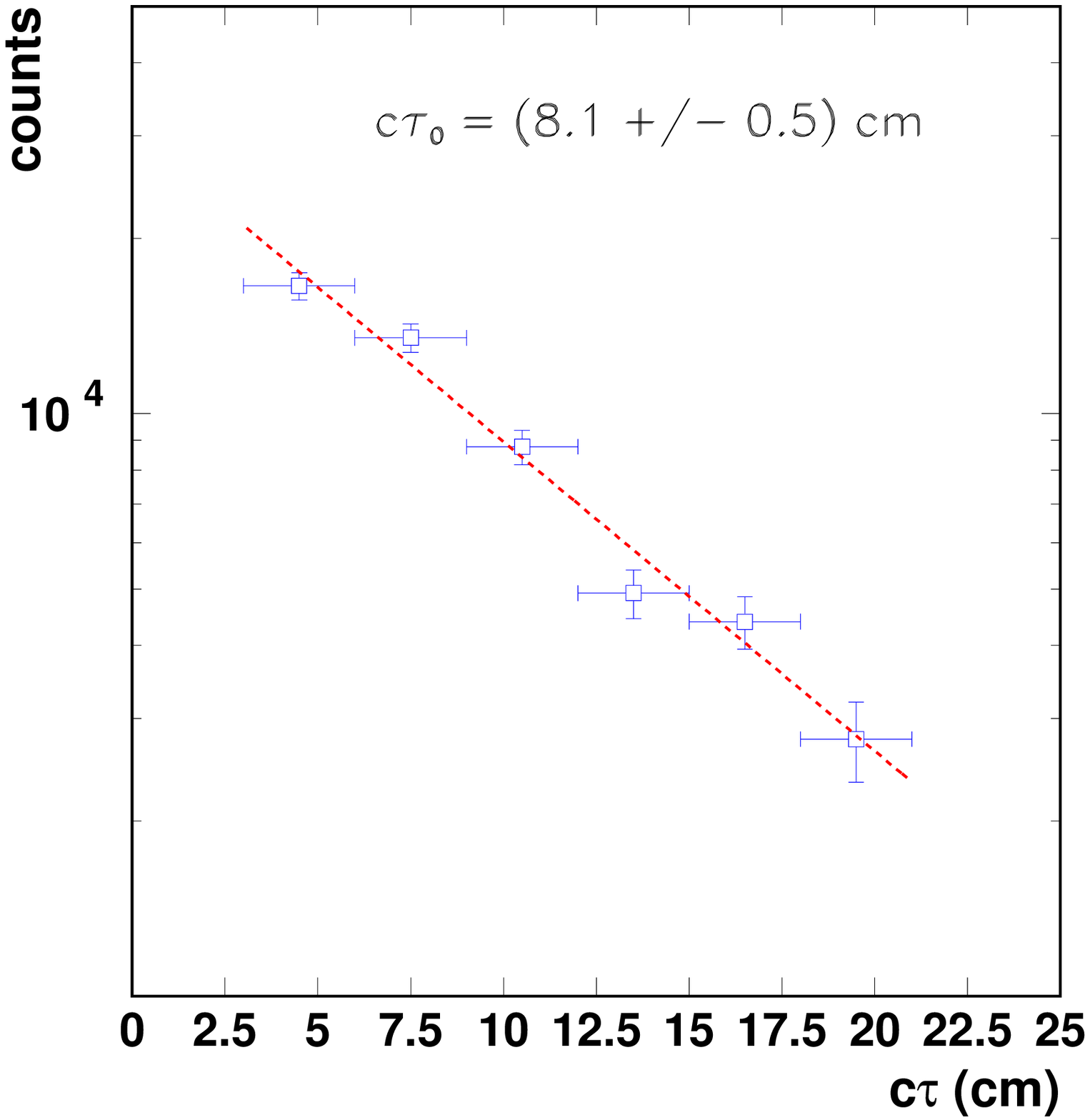}

\caption{Left: Invariant mass spectra for different extraction
    methods. On top are plotted the signal plus background (dashed
    line) and the normalized combinatorial background (solid
    line). The bottom shows the signal after background
    subtraction. \\
Right: Number of reconstructed $\Lambda$ as a function of the decay length.
}  
\label{consistency}
\end{figure}

{\bf Kinematical cuts:} The extraction of the strange-particle signal is
performed using geometrical and kinematical constraints. A fiducial cut
of 130 $\leq \theta \leq$ 240 mrad was placed on individual tracks. Due
to a loss in tracking efficiency at low momentum, a lower
transverse-momentum cut of 250 MeV/c was introduced for negatives. These
conditions affect the $\Lambda$ acceptance as shown in
Fig.~\ref{simulation}, middle. Additionally, a lower $p_t$ cut of 500
MeV/c for positives and an upper $p_t$ cut of 2.0 GeV/c for positives
and 0.6 GeV/c for negatives are introduced since they cost no signal and
reduce background. Also, only pairs with an opening angle greater 20
mrad are accepted. For a clean $\Lambda$ acceptance, we reject all
$\Lambda$ hyperons with $p_t \leq$ 0.9 GeV/c and $p_t \geq$ 2.5
GeV/c. There is also a fiducial cut in $\Lambda$ rapidity: $2.0 < y
<2.4$. Fig.~\ref{armenteros} shows an Armenteros plot of the data and
calculated lines for $\Lambda$ and K$_{\rm S} ^0$. In order to isolate
clean $\Lambda$ particles, additional cuts were made on $p_t^+$
selecting 10 - 120 MeV/c and on the Podolanski-Armenteros variable
$\alpha$ selecting values below 0.65, i.e.\ only $\Lambda$'s where the
proton is emitted backwards in the $\Lambda$ rest frame.

\section{Consistency tests:} In Fig.~\ref{consistency}, left, the
invariant mass spectrum is plotted for three different methods
analyzing the same events; (1) TPC in a stand-alone mode, (2) TPC plus
SDD veto and (3) TPC plus SDD veto plus vertex cut. One can notice a
loss in signal but a large improvement in the signal-to-background
ratio as successive cuts are introduced.  The $\Lambda$ signal,
obtained after subtraction of combinatorial background, is fitted with
a Gaussian. The fit parameters for all three methods are very similar,
the average of m$_{\Lambda}$ = 1.117 $\pm$ 0.001 GeV/c$^2$ is close to
the accepted value \cite{pdg} and the resolution is $\sigma
_{\Lambda}$ = 11.2 $\pm$ 0.4 MeV/c$^2$.

As an additional consistency test the acceptance corrected number of
reconstructed $\Lambda$'s is plotted as a function of the decay length in the
$\Lambda$ rest frame, as shown in the right part of Fig.~\ref{consistency}.
The superimposed solid line represents an exponential fit with a proper decay
length $c \tau_{0}$ = (8.1 $\pm$ 0.5) cm, in good agreement with the
accepted \cite{pdg} value of 7.89 cm.

\begin{figure}

 \includegraphics[width=68mm]{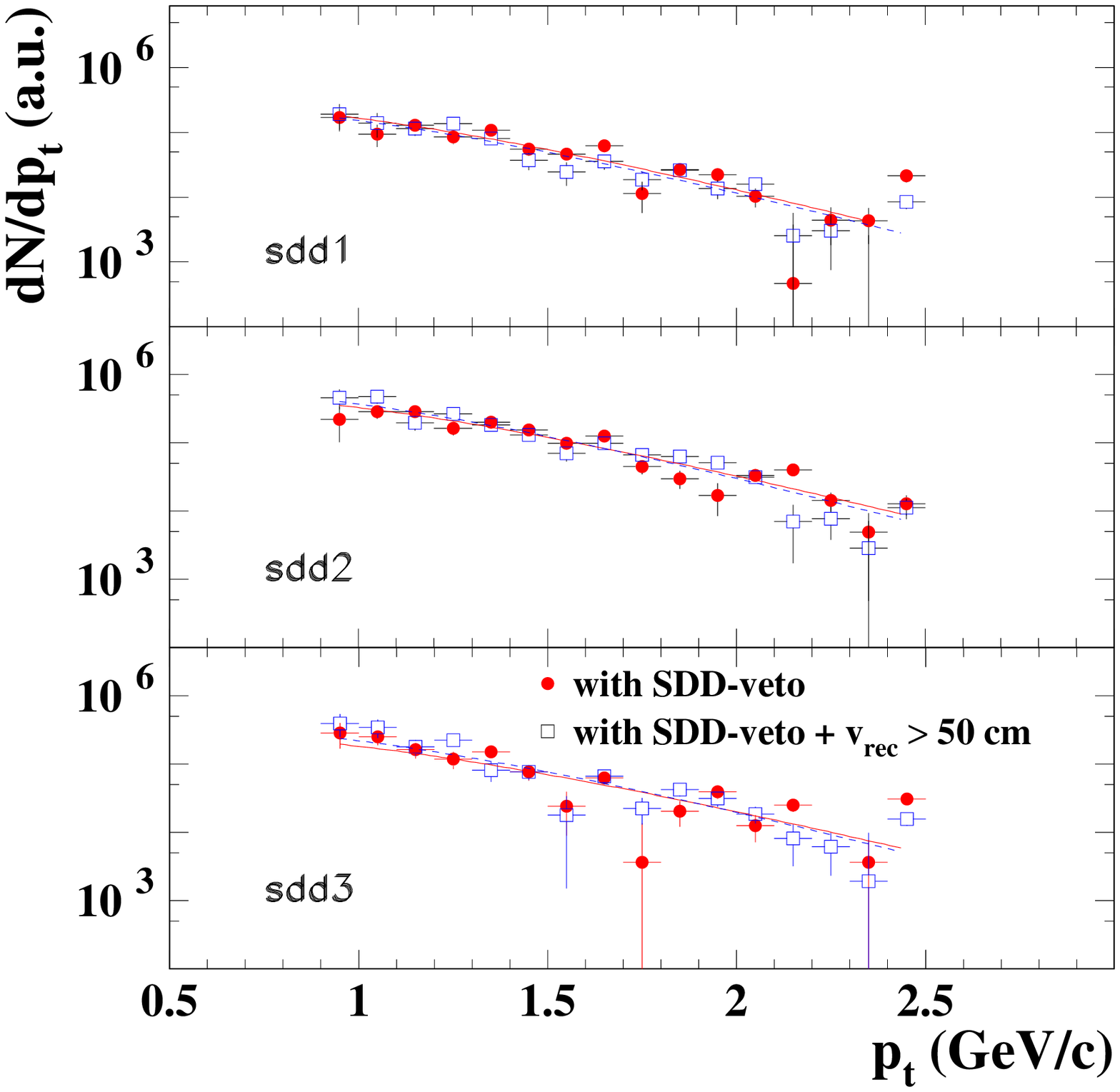}
\raisebox{2mm}{
 \includegraphics[width=65.5mm]{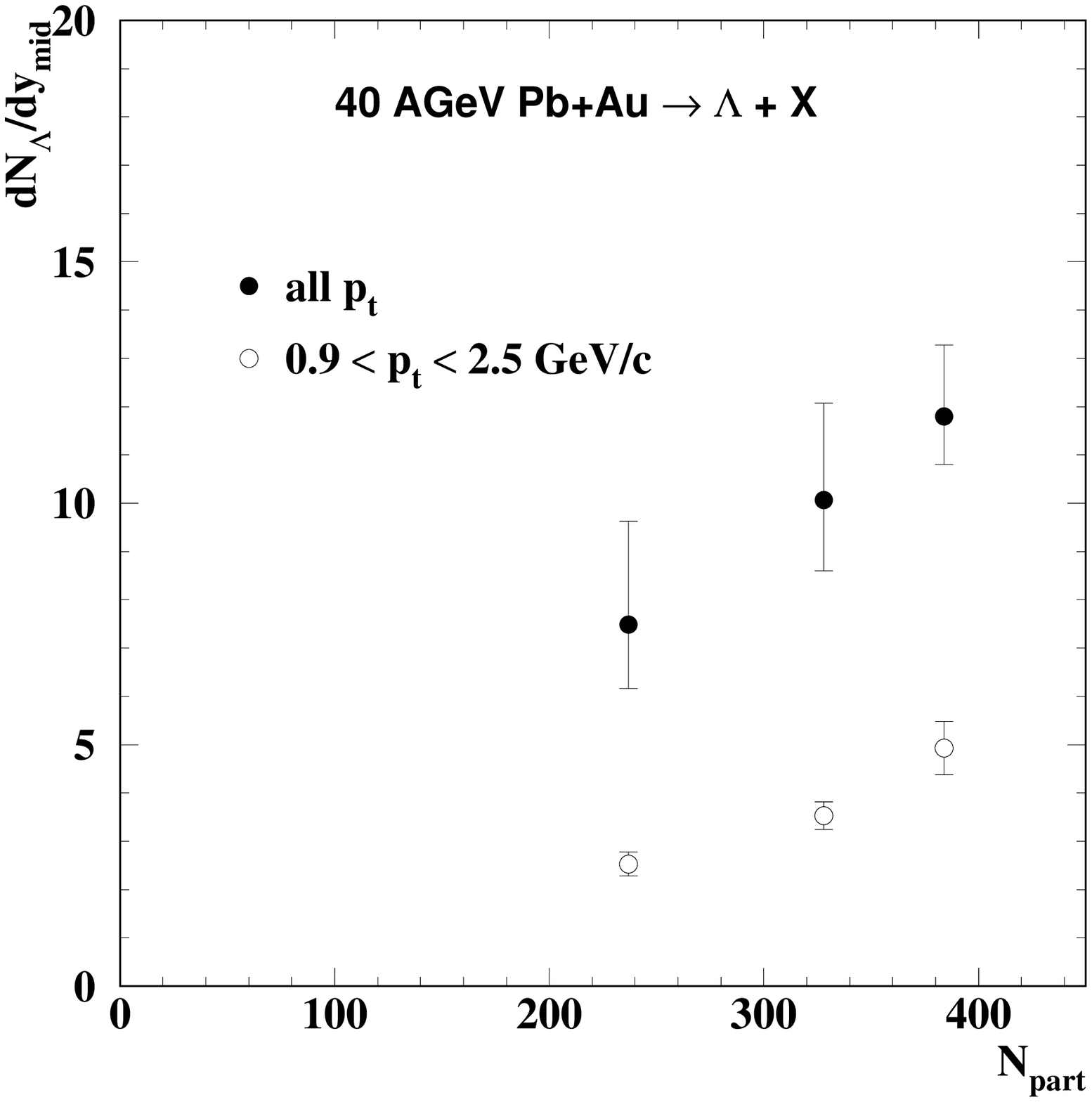}
}
     \caption{
Left: Transverse momentum spectra for $\Lambda$ shown
    separately for two different analysis methods and for the three
    centrality classes in the rapidity region 2 $<$ y$_{\Lambda}$ $<$
    2.4. All spectra are acceptance corrected but not normalized to
    the number of events.\\
Right: Midrapidity density dN$_{\Lambda}$/dy$_{\rm {mid}}$ 
as a function of the centrality of the collision. The plot shows data
for the $p_t$ 
range of our measurement as well a an extrapolation to all $p_t$ (see text).
} 
\label{results1}
\end{figure}

\section{Centrality classes}

The event sample was divided into three centrality classes
according to the number of SDD1/SDD2 tracks as indicated
in Tab.~\ref{tab-mult}. Also indicated there are the corresponding
fractions of the geometric cross section and the number of
participants $N_{part}$ evaluated using the UrQMD model \cite{urqmd}.

\begin{table}
 \begin{center}
 \begin{tabular}{|c|c|c|c|} \hline
 mult. class & sdd1 & sdd2 & sdd3 \\ \hline
 SDD-tracks & 100 - 200 & 200 - 300 & 300 - 400 \\ \hline
 events ($10^6$) & 0.8 & 1.3 & 0.7 \\ \hline
 $\sigma / \sigma _{\rm {geo}}$ (\%) & 15 - 36.7 & 4.8 - 15 & $<$ 4.8 \\ \hline
 mean $N_{\rm {part}}$ & 237 & 328 & 384 \\ \hline
 \end{tabular}
 \end{center}
 \caption{Definition of the three centrality classes}
 \label{tab-mult}
\end{table}

\section{Results}

The experimental data were corrected for the geometrical acceptance
and reconstruction efficiency. In Fig.~\ref{results1}, left, the
corrected but un-normalized $\Lambda$ spectra are presented in different
multiplicity bins for two different analysis methods. The invariant
multiplicity is fitted with an exponential distribution
${dN}/{p_t dp_t} ~ \propto ~ \exp ( - {m_t}/{T_{\Lambda}})$,
also shown in Fig.~\ref{results1}.

\begin{figure}[tbh]
  \includegraphics[width=65mm]{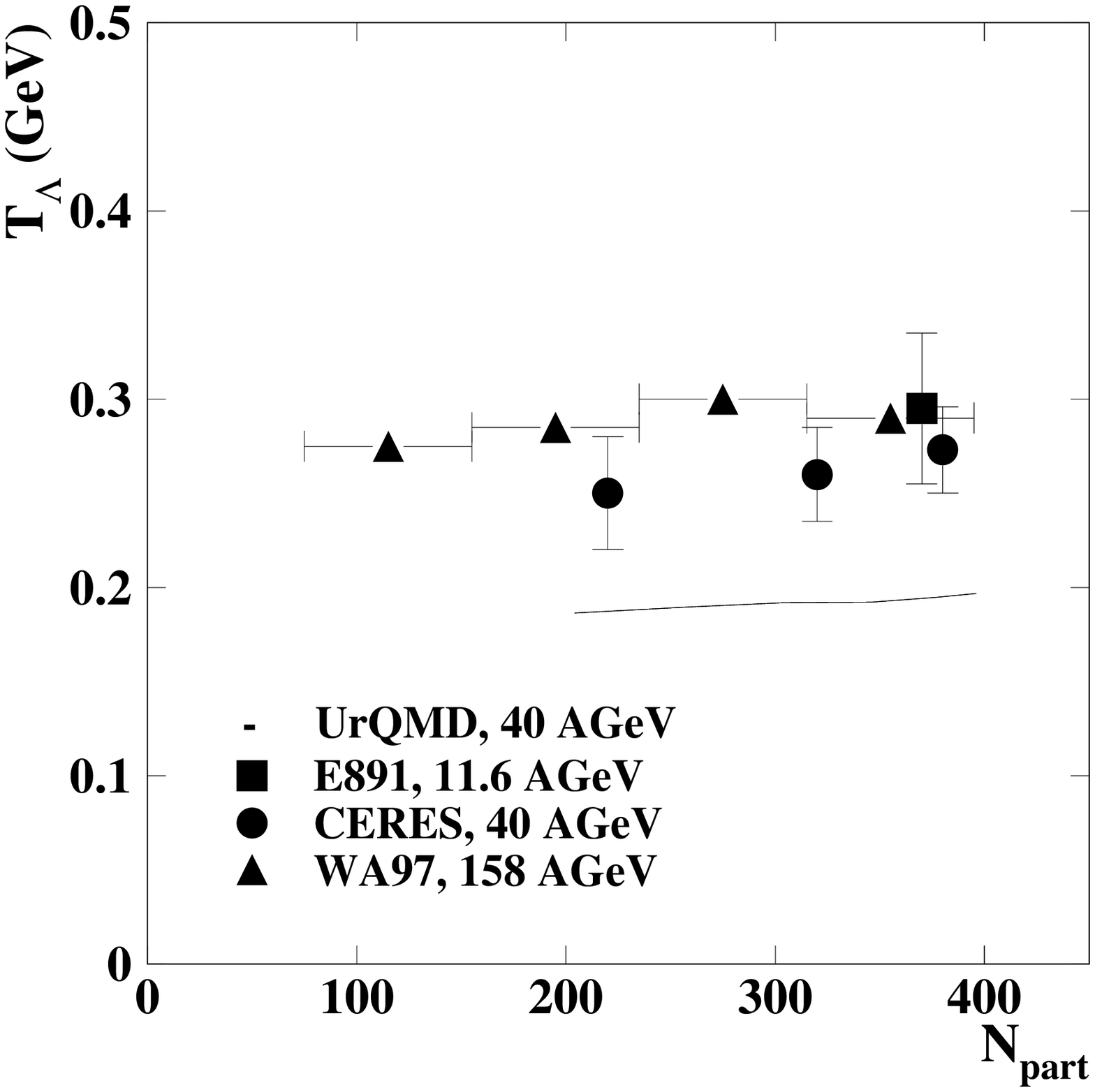}
    \includegraphics[width=65mm]{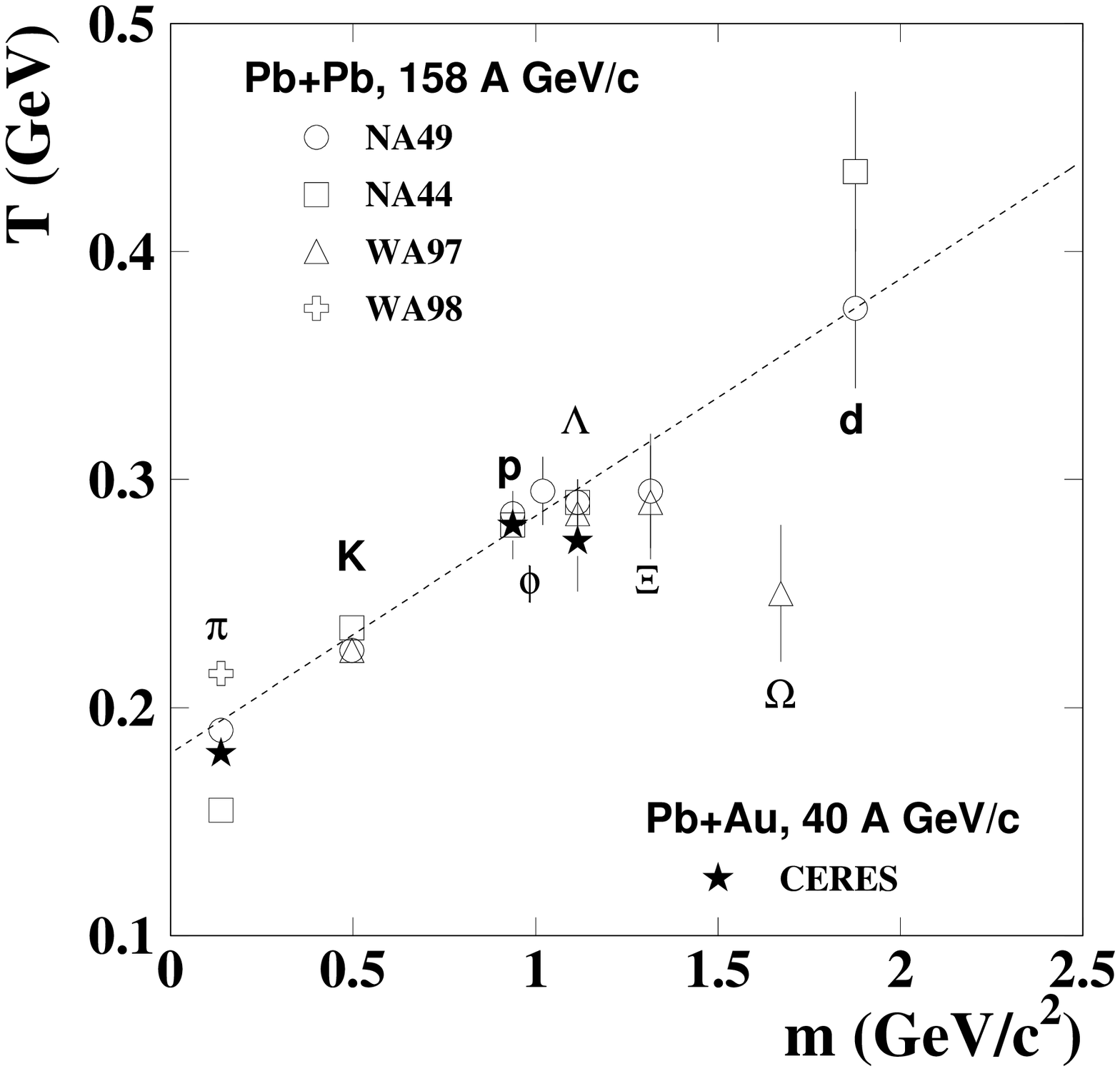}

\caption{Left: Dependence of the inverse slope
    parameter T$_{\Lambda}$ on the centrality of the collision. In
addition to the present data are shown data from AGS and full energy
SPS as well as a calculation in the UrQMD model (references see
text). \\ 
Right:
    Mass dependence of inverse slope for the most central
    collisions. The CERES results for 40 A\,GeV/c (present data and
    \cite{Harry}) are shown together with the systematics for full SPS
    energy \cite{INPC}
} 
\label{results2}
\end{figure}

The dependence of the inverse slope parameter T$_{\Lambda}$ on the number of
participants N$_{\rm {part}}$ is presented in
Fig.~\ref{results2}, left.  The measured values of T$_{\Lambda}$
exhibit a slight but not very significant increase as a function of
centrality. As shown the slopes are close to those observed at top AGS and SPS
energies. The measured values are not well reproduced by the UrQMD model
(line in Fig.~\ref{results2}), as also observed for the slopes of the
proton spectra \cite{Harry}. The mass dependence of the inverse slope
is presented in Fig.~\ref{results2}, right. Within errors, our present
data agree with the systematic behavior of measurements at top SPS
energies \cite{INPC}. The large differences between the slopes 
\mbox{(T$_{\rm h ^-}$ = 176 $\pm$ 5 MeV, T$_{(+)-(-)}$ = 278
$\pm$ 12 MeV and T$_{\Lambda}$ = 273 $\pm$ 20 MeV)} indicate the
presence of a strong radial flow also at 40 A\,GeV beam energy
\cite{Harry}.

\begin{figure}[tbh]
  \includegraphics[width=60mm]{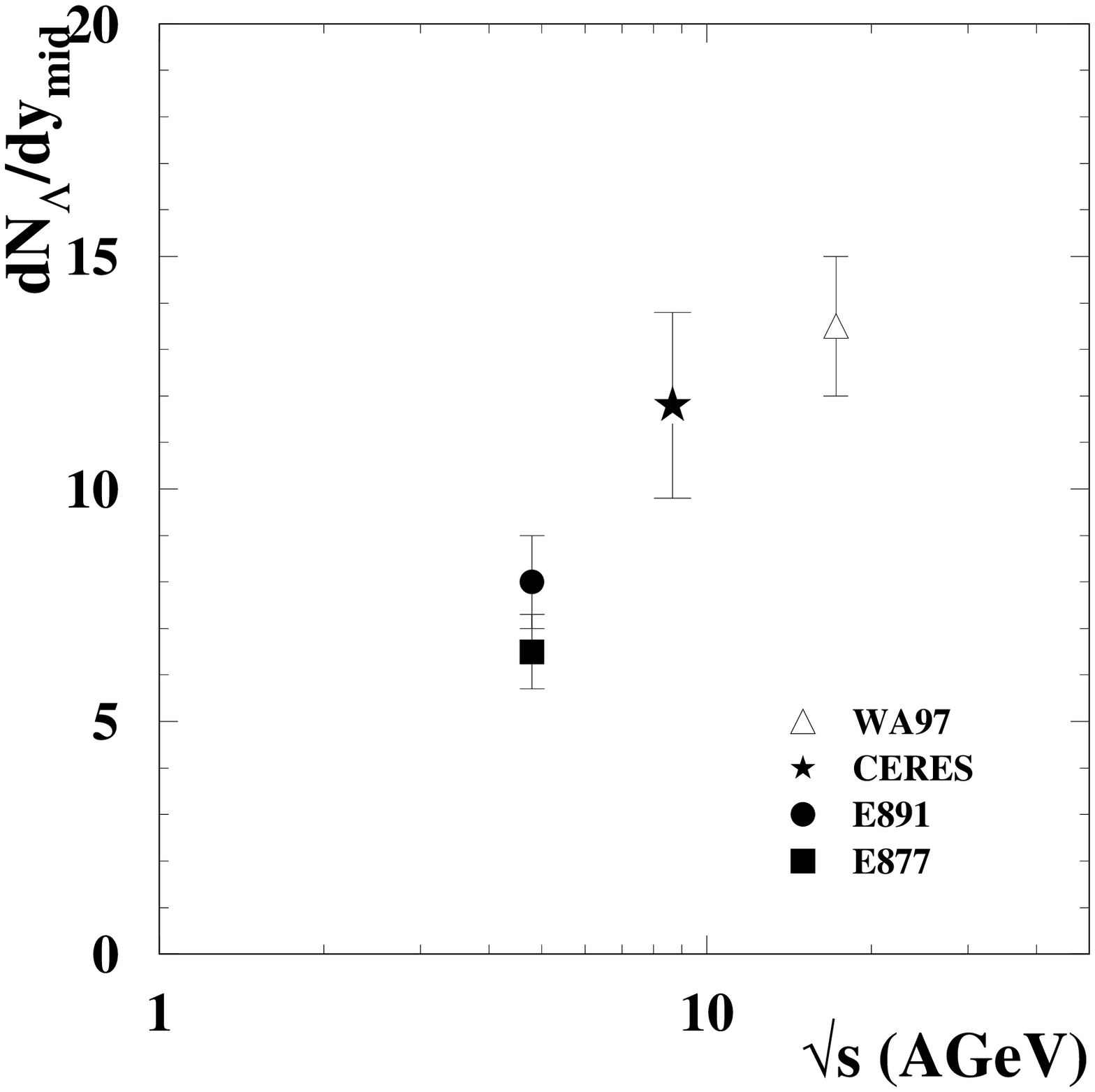}
  \includegraphics[width=60mm]{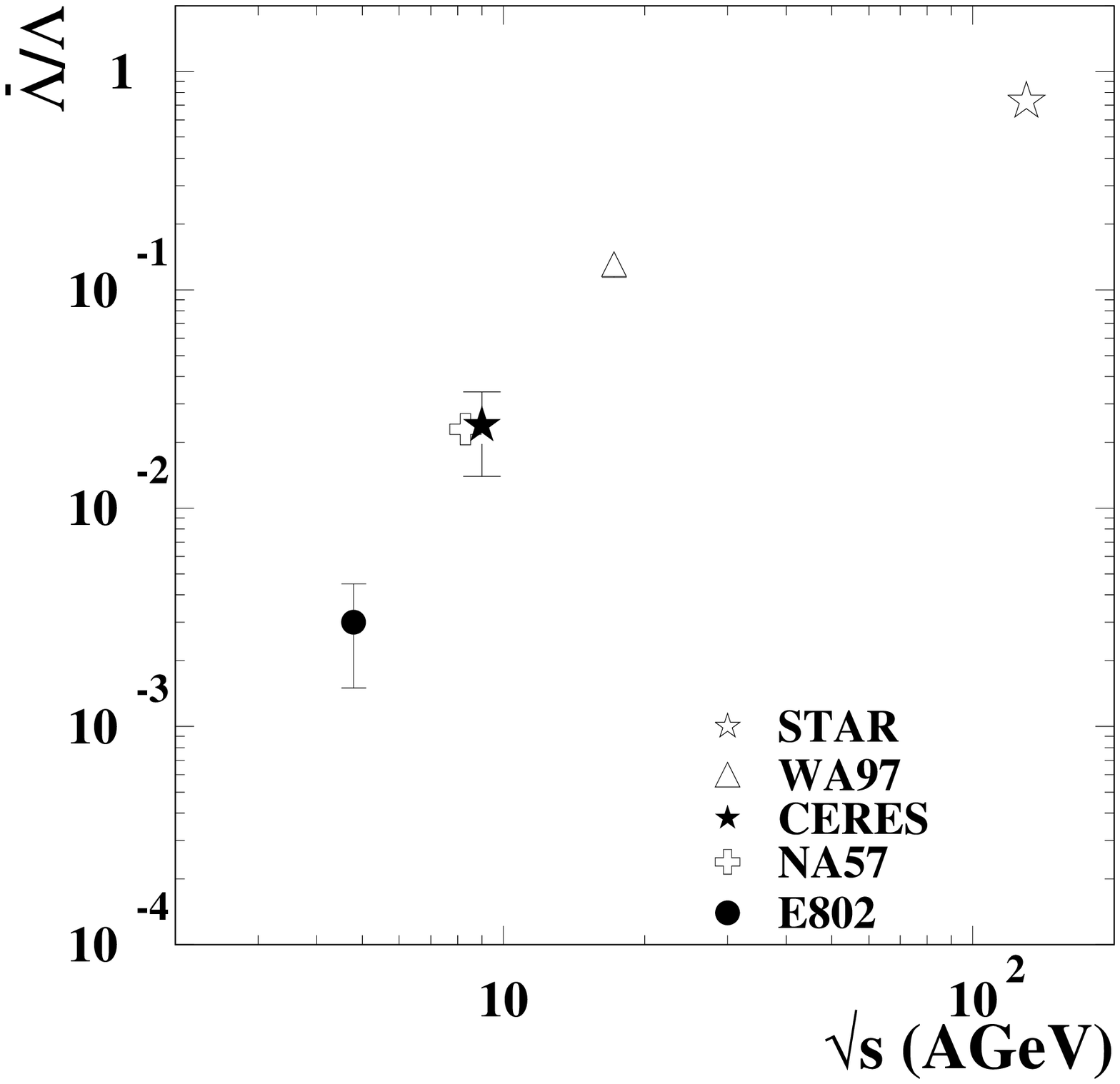}
     \caption{
Left: dN$_{\Lambda}$/dy$_{\rm {mid}}$ as a function of the beam energy
 $\sqrt{\rm s}$ (data for AGS and full energy SPS energies in 
\cite{lam891,lam877,lamwa97}).\\
 Right: $\bar{\Lambda}/\Lambda$ ratio as a 
function of the beam energy $\sqrt{\rm s}$ (other data from
\cite{802,NA57,lamwa97,STAR}).}
    \label{results3}
\end{figure}

The midrapidity density dN$_{\Lambda}$/dy$_{\rm {mid}}$ has been
obtained by extrapolating the transverse momentum spectra to p$_{\rm t}$
= 0 using the functional form given above. On the right side of
Fig.~\ref{results1} the rapidity densities in the measured
p$_{\rm t}$ interval as well as extrapolated to p$_{\rm t}$ = 0 are
presented as a function of centrality. A continuous rise with centrality
is observed. In our most central multiplicity class we determine a total
midrapidity density dN$_{\Lambda}$/dy$_{\rm {mid}}$ = 11.8 $\pm$
2. A somewhat larger value of about 15 is reported by the NA49
collaboration \cite{NA49}.
Putting our present data into context with measurements at lower and
higher beam energy, a continously rising behavior is seen as a function
of $\sqrt{\rm s}$ (see Fig.~\ref{results3}, left).

 \begin{figure}[b]
  \includegraphics[width=70mm]{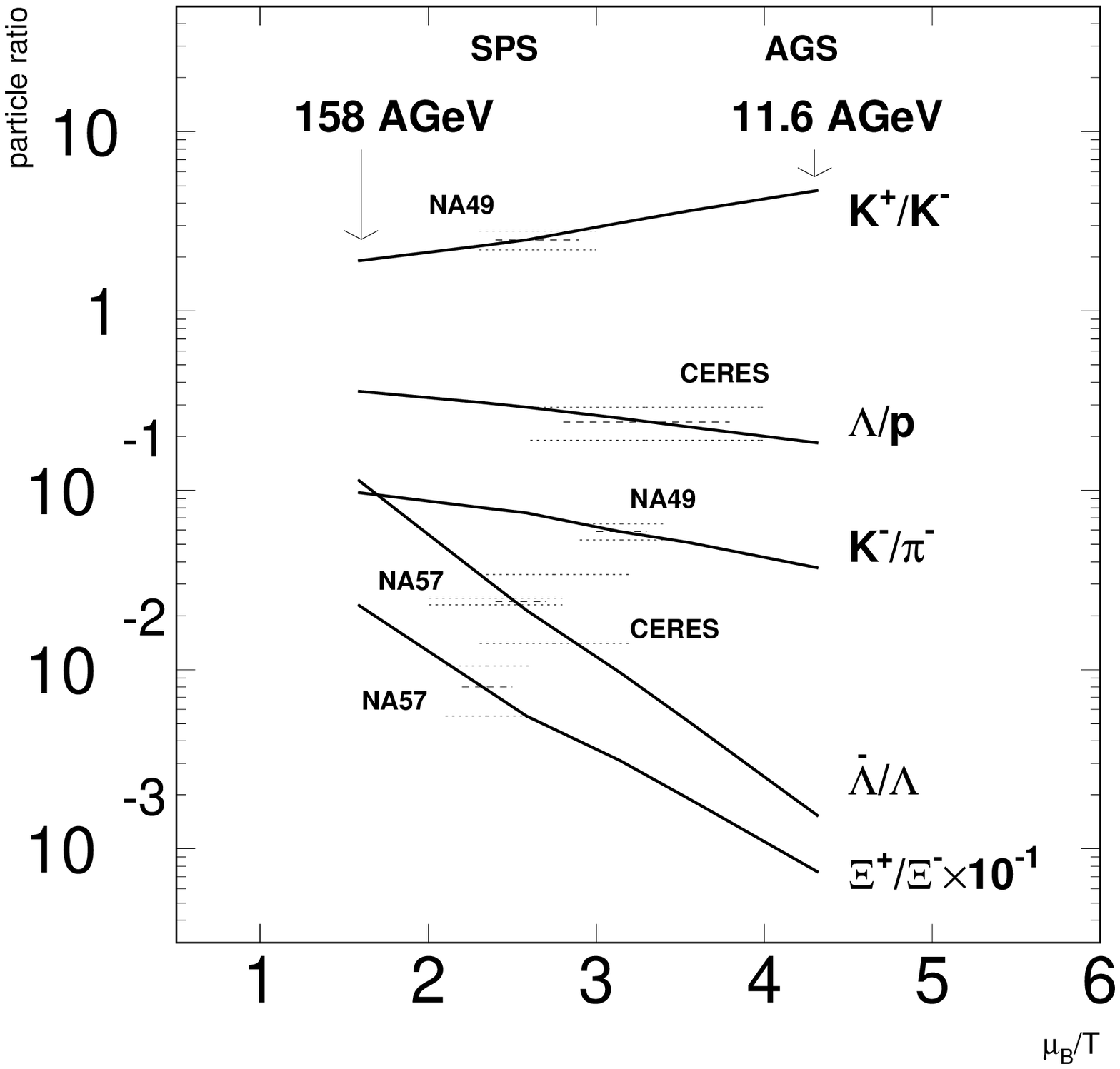}
\raisebox{1.5mm}{
    \includegraphics[width=61.5mm]{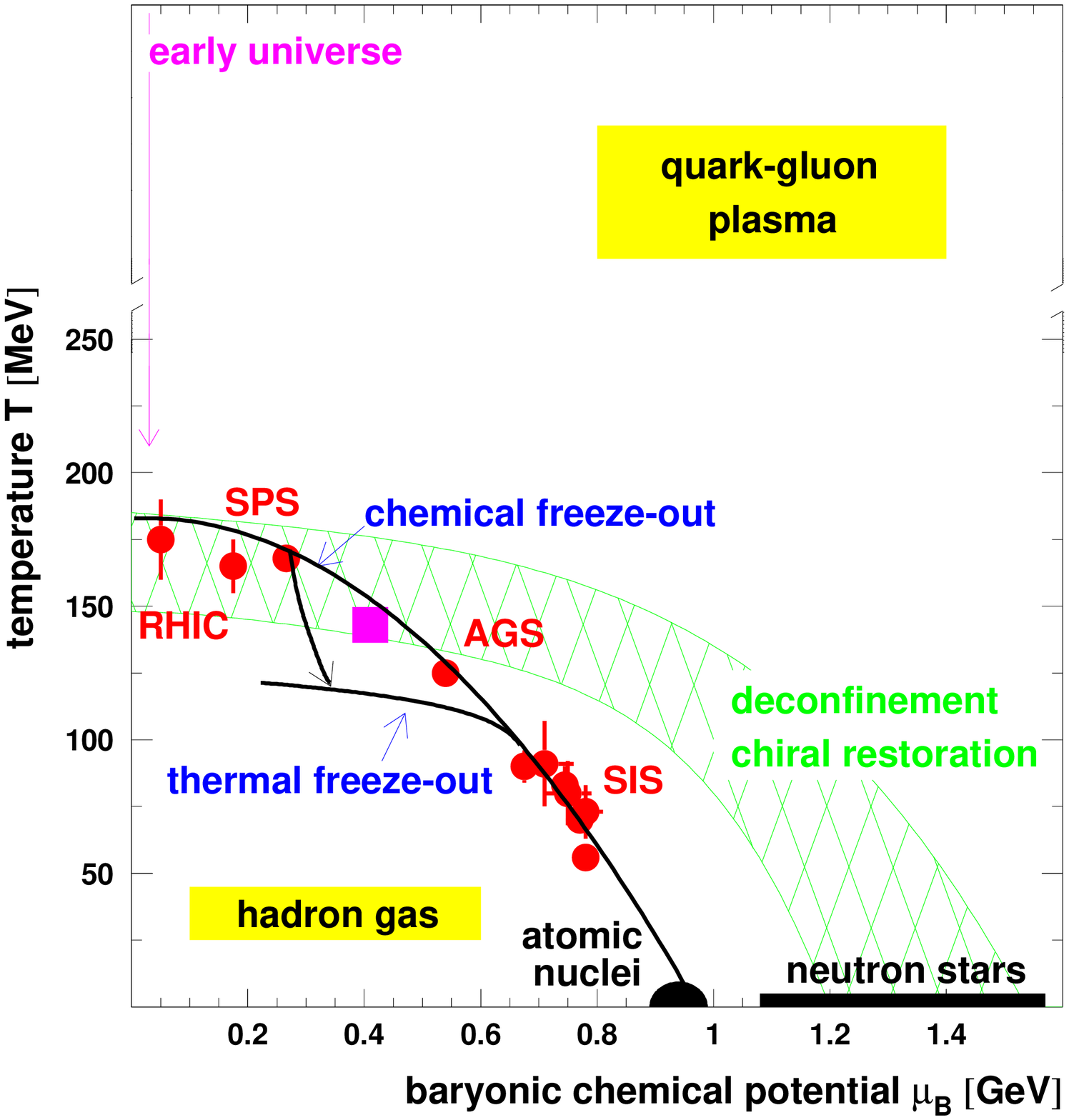}}
  \caption{Left: Comparison of data \cite{ratios,NA57} and particle ratios 
    computed utilizing a thermal model \cite{Stachel}.\\ 
  Right: Phase diagram of nuclear matter \cite{plott} with the present
data point for 40 A\,GeV (square symbol).}
    \label{ratios}
\end{figure}

We have also extracted the signal for the $\bar{\Lambda}$. But due to the low
statistics it is not possible to show transverse momentum spectra for
$\bar{\Lambda}$. Since all corrections are symmetric for particle and
antiparticle, they cancel in the ratio and we obtain R$_{\Lambda}$ =
$\bar{\Lambda}/\Lambda$ = 0.024 $\pm$ 0.010. The dominant error is
statistical. Our result is in good agreement with the value recently reported
by NA57 \cite{NA57}, albeit with a much larger statistical error for our
present result.  The $\bar{\Lambda}/\Lambda$ ratio as a function of $\sqrt{\rm
  s}$ (Fig.~\ref{results3}, right) shows the expected steep rise
with increasing beam energies; for a net-baryon-free region the
antiparticle/particle ratio is expected to be one and at RHIC energies this
values is approached.

Information about the chemical freeze-out parameters of the expanding
system, the temperature T and the baryonic chemical potential $\mu _{\rm
B}$, can be obtained from the ratios of particle yields. Utilizing the
thermal model of \cite{Stachel} we have calculated several particle
abundancies corresponding to the phenomenological condition
\cite{Cleymans} of fixed average energy per hadron of 1 GeV between top
AGS and SPS energies, characterized by a parameterization in $\mu_{\rm
B}$/T.  The comparison of computed and measured particle ratios is
presented in Fig.~\ref{ratios}, left. Taking the different measured
particle ratios into account we could extract the mean ratio $\mu _{\rm
B}$/T = 2.8$\pm 0.3$. Following the Cleymans-Redlich freeze-out curve
\cite{Cleymans}, this corresponds to values of $\mu _{\rm B}$ = 410
$\pm$ 30 MeV and T = 143 $\pm$ 5 MeV. This result is shown in the phase
diagram Fig.~\ref{ratios}, right. The chemical freeze-out point at 40
A\,GeV is near the phase boundary and falls into the systematics between
top AGS and SPS energies.

\end{document}